\documentclass[journals]{IEEEtran}
\IEEEoverridecommandlockouts
\usepackage{cite}

\usepackage{amsmath,amssymb,amsfonts}
\usepackage{amsthm}
\usepackage{algorithmic}
\usepackage[ruled,noend,vlined,linesnumbered]{algorithm2e}
\usepackage{graphicx}
\usepackage{textcomp}
\usepackage{xcolor}
\usepackage{authblk}

\usepackage{booktabs}
\usepackage{multirow}
\newcommand{\D}{\mathcal{D}}
\newcommand{\E}{\mathbb{E}}
\newcommand{\R}{\mathbb{R}}

\def\BibTeX{{\rm B\kern-.05em{\sc i\kern-.025em b}\kern-.08em
    T\kern-.1667em\lower.7ex\hbox{E}\kern-.125emX}}

\newtheorem{definition}{Definition}

\newcommand{\bheading}[1]{{\vspace{2pt}\noindent{\textbf{#1}}\hspace{2pt}}}

\begin{document}

\title{Robust and Privacy-Preserving Collaborative Learning: A Comprehensive Survey\\
}

\author{%
  \IEEEauthorblockN{%
  Shangwei Guo\IEEEauthorrefmark{1},
  Xu Zhang\IEEEauthorrefmark{1},
  Fei Yang\IEEEauthorrefmark{2}\textsuperscript{\textsection},
  Tianwei Zhang\IEEEauthorrefmark{3},
  Yan Gan\IEEEauthorrefmark{1},
  Tao Xiang\IEEEauthorrefmark{1}, and
  Yang Liu\IEEEauthorrefmark{3}
  }%

  \IEEEauthorblockA{\IEEEauthorrefmark{1}\textit{College of Computer Science, Chongqing University, Chongqing, China}}

  \IEEEauthorblockA{\IEEEauthorrefmark{2}\textit{Zhejiang Lab, Hangzhou, China}}

  \IEEEauthorblockA{\IEEEauthorrefmark{3}\textit{School of Computer Science and Engineering, Nanyang Technological University, Singapore}}
}
\maketitle
\begingroup\renewcommand\thefootnote{\textsection}
\footnotetext{Fei Yang is the corresponding author (email: yangf@zhejianglab.com).}
\endgroup


\begin{abstract}
With the rapid demand of data and computational resources in deep learning systems, a growing number of algorithms to utilize collaborative machine learning techniques, for example, federated learning, to train a shared deep model across multiple participants. It could effectively take advantage of resource of each participant and obtain a more powerful learning system. However, integrity and privacy threats in such systems have greatly obstructed the applications of collaborative learning. And a large amount of works have been proposed to maintain the model integrity and mitigate the privacy leakage of training data during the training phase for different collaborate learning systems. Compared with existing surveys that mainly focus on one specific collaborate learning system, this survey aims to provide a systematic and comprehensive review of security and privacy researches in collaborative learning. Our survey first provides the system overview of collaborative learning, followed by an brief introduction of integrity and privacy threats. In an organized way, we then detail the existing integrity and privacy attacks as well as their defenses. We also list some open problems in this area and opensource the related papers on GitHub: https://github.com/csl-cqu/awesome-secure-collebrative-learning-papers.
\end{abstract}


\section{Introduction}\label{sec:intro}
Deep learning (DL) has demonstrated its tremendous success in multiple fields including computer vision, natural language processing, bioinformatics, and board game programs. DL systems adopt deep neural networks (DNNs) to improve automatically through experience on huge training datasets \cite{krizhevsky2012imagenet, szegedy2015going, graves2013speech, devlin2018bert}. To efficiently train a DL model, a learning system mainly relies on two components: a large number of high-quality training samples and high-performance GPUs. However, the training datasets and GPUs may be distributed at different parties due to various reasons. Consider the following two examples \cite{litjens2017survey,gawali2021comparison,hard2018federated}:
\newtheorem*{example1}{Medical Image Classification}
\newtheorem*{example2}{Mobile Keyboard Prediction}

\begin{example1}
    A hospital wants to learn a lung cancer detector model to assist its doctors in identifying lung cancer patients from their computed tomography (CT) images. Because the hospital has only received a limited number of lung cancer patients, learning a highly accurate model is difficult for it. To guarantee the accuracy of the diagnosis, the hospital unites with other hospitals to collaboratively learn a shared model together. All hospitals need to keep their CT images locally by considering the privacy of patients.
\end{example1}
\begin{example2}
     As users increasingly shift to mobile devices, Gboard, the Google keyboard, indents to provide reliable and fast mobile input methods such as next-word predictions. Although publicly available datasets can be used for such task, the distribution of such datasets often does not match that of users. Thus, Gboard requires user-generated texts for better performance without causing users to be uncomfortable with the collection and remote storage of their personal data.
\end{example2}

Collaborative learning has recently been popular as a promising solution for such application scenarios \cite{dean2012large, peteiro2013survey,leroy2019federated, he2020group, zheng2020federated,lim2020federated,aledhari2020federated}. Specifically, collaborative learning allows two or more participants to collaboratively train a shared global DL model while keeping their training datasets locally. Each participant trains the shared model on his own training data and exchanges and updates model parameters with other participants. Collaborative learning can improve the training speed and the performance of the shared model while protecting privacy of the participants’ training datasets. Thus, it is a promising technique for the scenarios where the training data is sensitive (e.g., medical records, personally identifiable information, etc.). Several learning architectures have been proposed for collaborative learning: with and without a central server, with different ways of aggregating models, etc \cite{li2014scaling, moritz2015sparknet, liu2019decentralized, sun2021stability, sahu2018convergence, reddi2020adaptive, wang2020tackling,lu2021optimal}. An important branch of collaborative learning is federated learning \cite{li2021survey} that enables mobile phones to collaboratively learn a shared prediction model while keeping all the training data on device, decoupling the ability to do machine learning from the need to store the data in the cloud.

Although each participant stores his training dataset locally and only shares the updates of the global model at each iteration, adversaries can also conduct attacks to break model integrity and data privacy during the training process \cite{guerraoui2018hidden, bhagoji2019analyzing, zhu2019deep,zhang2018survey}. One of the most severe threats is the model integrity that can be easily compromised when some of participants are not trustworthy \cite{NIPS2017_f4b9ec30,guo2021byzantine}. For example, malicious participants poison their training datasets with some carefully crafted malicious triggers. Then, at each iteration, they generate malicious updates with the triggers and gradually inject such triggers as backdoors into the global model by sharing the malicious updates to earn extra profit or increase their advantages \cite{bagdasaryan2020backdoor, wang2020attack}. Adversaries can also disguise as participants to join the collaborative learning process and destroy the learning process by sending malicious updates to their neighborhoods or parameter servers \cite{munoz2017towards, bhagoji2019analyzing,baruch2019little}. Blanchard et al. \cite{NIPS2017_f4b9ec30} and Guo et al. \cite{guo2021byzantine} show that only one malicious participant can control the whole collaborative learning process.

Other than the model integrity threats, another crucial challenge is to protect the data privacy of each participant. It have been demonstrated that although participants do not share the raw training samples with others, the shared updates are generated from the samples and also leak information about the training datasets indirectly. For instance, Melis et al. \cite{melis2019exploiting} found that one can capture the membership and unintended feature leakage from the shared gradients during the training process. More seriously, Zhu et al. \cite{zhu2019deep} proposed an optimization method that can reconstruct training samples from the corresponding updates.

To address the above integrity and privacy threats, many methods are proposed to defend these attacks \cite{NIPS2017_f4b9ec30, cao2019distributed, guerraoui2018hidden, munoz2019byzantine, pan2020justinian, shejwalkar2021manipulating, xie2019zeno, xie2020zeno++, yin2018byzantine, tran2018spectral,chen2018detecting,chan2019poison,chou2018sentinet,gao2019strip,truong2020systematic,ma2019nic,liu2019abs,wang2019neural,huang2019neuroninspect,chen2019deepinspect,ma2019nic,sun2019can,liu2020backdoor,zhao2020shielding,liu2019abs,zhu2019deep,2020Defending,chaudhuri2011differentially,abadi2016deep,zhang2018improving,li2018differentially,yu2019differentially,jayaraman19evaluatiing,aono2016scalable, kim2018secure,bonawitz2017practical,li2020privacy}. For instance, to achieve byzantine-resilient collaborative learning, Blanchard et al. \cite{NIPS2017_f4b9ec30} use statistic tools to inspect the updates of participants at each iteration and abandon potential malicious updates when aggregating updates. In terms of privacy protection, Gao et a. \cite{gao2021privacy} proposed to search privacy-preserving transformation functions and pre-process the training samples with such functions to defend reconstruction attacks as well as preserving the accuracy of the trained DL models. Several defenses \cite{ma2022privacy,grama2020robust,naseri2020toward,qi2021privacy,liu2021privacy} also proposed robust and privacy-preserving defenses to defend both integrity and privacy threats.

Several survey works \cite{lyu2020privacy, lyu2020threats, mothukuri2021survey, zhang2018survey,liu2019survey,vepakomma2018no,kairouz2019advances,enthoven2020overview,yang2020adversary} have also summarized some of the threats and defenses in collaborative learning. However, they have certain drawbacks. First, most of them only consider some specific branches of collaborative learning and lack systematic and comprehensive analysis towards other collaborative learning systems. . For example, Second, several surveys \cite{lyu2020privacy, lyu2020threats,enthoven2020overview} mainly target on the threats and defenses in federated learning. Vepakommacite et al. {vepakomma2018no} summarize the privacy problems and defenses in distributed learning systems. Second, existing surveys do not focus on the training process of collaborative learning systems (the most important stage) and selectively introduce existing threats and defenses, which makes them unable to summarize state-of-the-art methods well.

In this paper, we focus on the integrity and privacy attacks and defenses during the training process of collaborative learning and present a comprehensive survey of the state-of-the-art solutions. Specifically, we systematically introduce different types of collaborative learning systems from various perspectives (Section \ref{sec:system}). Then, we summarize summarizes the privacy and integrity threats in collaborative learning in Section \ref{sec:threat}. One the one hand, we exhibits existing attacks and the corresponding defenses in Section \ref{sec:integrity_attack} and \ref{sec:integrity_defense}, respectively. On the other hand, we shows the state-of-the-art integrity privacy attacks and the corresponding defenses in \ref{sec:integrity_defense}  Section \ref{sec:privacy_attack}, respectively. We summarize hybrid defense methods to achieve robust and privacy-preserving collaborative learning and adversarial training algorithms to improve the robust of model inference.
We illustrate some open problems and future solutions in collaborative learning in  Section \ref{sec:open_problem}, followed by Section \ref{sec:conclusion} that concludes this paper.
We also opensource the paper list of the attack and defense methods on GitHub: https://github.com/csl-cqu/awesome-secure-collebrative-learning-papers.

\section{System Overview}\label{sec:system}

\subsection{Machine Learning Model}
In general, machine learning is categorized as supervised learning \cite{caruana2006empirical}, unsupervised learning \cite{figueiredo2002unsupervised} and reinforcement learning \cite{kaelbling1996reinforcement,chen2021stealing}. In this survey, machine learning is mostly referred to as supervised learning. Supervised learning is the process of optimizing a function from a set of labeled samples such that, given a sample, the function would calculate an approximation of the label.

We use a \emph{dataset} $\D$ to denote a probability distribution of data; $z {\sim}\D$ denotes a random sampled variable $z$ from $\D$, and $\E_{z{\sim}\D}[f(\xi)]$ denotes the expected value of $f(\xi)$ for a random variable $\xi$. For a deep learning model, we use $w\in\R^d$ be the $d$-dimensional parameter vector to estimate the model; $L_{\D}(f)$ be the loss calculated by $f$ on dataset $\D$;  $l$ be the loss function of an individual sample. Therefore, we abstract the machine learning task by the following optimization problem:
\begin{equation}\label{equ:optimization}
    w^{*}=\underset{w\in\R^d}{\mathit{argmin}}{L_{\D}}(f_w)=\underset{w\in\R^d}{\mathit{argmin}}\underset{\xi\sim\D}{\E}[l(w,\xi)]
\enskip.
\end{equation}

There are many different approaches \cite{battiti1992first} to minimize the loss function, such as gradient descent, second-order methods, evolutionary algorithms, etc. In machine learning, optimization is majored performed via gradient descent. By using randomly sampled data in each iteration, we can apply \emph{Stochastic Gradient Descent} (SGD) \cite{goyal2017accurate} to optimize Eq.~\ref{equ:optimization}.

\begin{figure*}[t]
	\centering
	\includegraphics[width=0.8\textwidth]{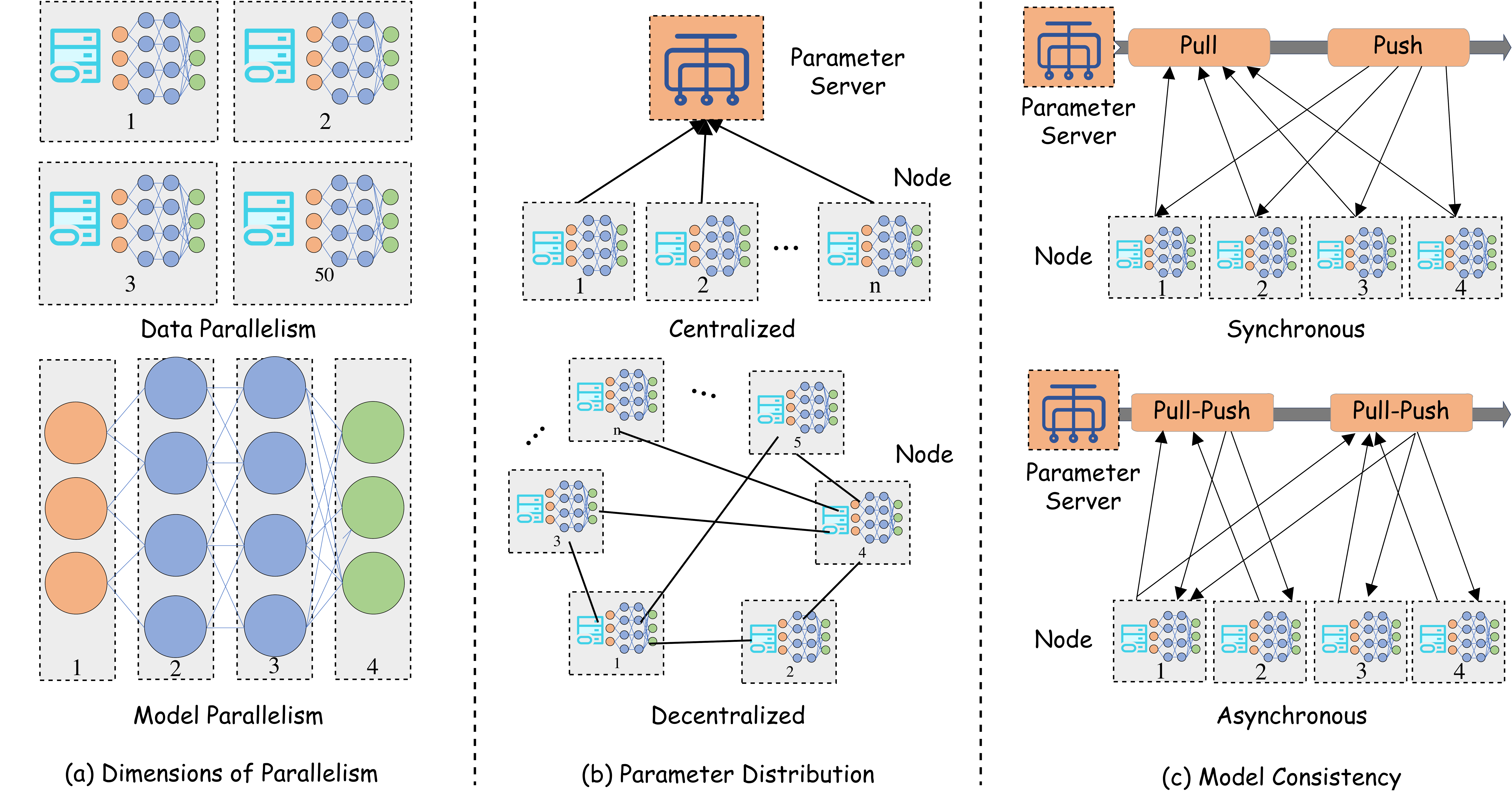}
	\caption{System Overveiw}
	\label{fig:MLStructure}
\end{figure*}


\subsection{Dimensions of Parallelism}
Machine learning is growing rapidly in the recent decade due to the growth of the sizes of models and datasets. Machine learning algorithms improved a lot thanks to the more and more complicated models, as well as larger and larger datasets. Therefore, parallelism is introduced to provide the scalability of machine learning algorithms. As illustrated in Fig.~\ref{fig:MLStructure}, parallel training allows users to partition the data and computation tasks onto multiple computational resources such as cores and devices. We introduce four prominent partitioning strategies, categorized by the dimension of parallelism, which are data parallelism, model parallelism, pipelining and hybrid parallelism.

\subsubsection{Data Parallelism}
For data parallelism \cite{KrizhevskySH12}, as shown in the top figure of Fig.~\ref{fig:MLStructure} (a), the approach is to partition the samples from the dataset among multiple computational resources (cores or devices). This method is the dominant distributed deep neural network training strategy.

\subsubsection{Model Parallelism}
Data parallelism (the bottom figure of Fig.~\ref{fig:MLStructure} (a)) sometimes suffers from very large models since the memory required to store parameters and activations and the time to synchronize parameters makes data parallelism impossible or inefficient. Model parallelism \cite{DeanCMCDLMRSTYN12} is introduced to solve the above issue. The strategy of model parallelism is to split the model into multiple computational resources. It divides the computational tasks according to the neurons in each layer. Moreover, the sample minibatch is copied to all processors and different parts of the model are computed on different processors.

\subsubsection{Pipelining}
Pipelining in machine learning can either refer to overlapping computations between layers or partitioning the DNN according to depth and assigning layers to specific processors. Therefore, pipelining is both a form of data parallelism as samples are processed by the network in parallel, and a form of model parallelism as models are partitioned by layers.

The pipelining strategy can be used to overlap the procedure of forward evaluation, backpropagation and weight updates. This approach minimizes the processor idle time. Meanwhile, pipelining could also be treated as partitioning the layers, each processor handles a fixed layer and the data flow is determined throughout the whole procedure.

\subsubsection{Hybrid Parallelism}
Hybrid parallelism combines multiple parallelism schemes. For instance, in AlexNet, a successful approach is to apply data parallelism to the convolutional layer where most computations are performed, and model parallelism to the fully connected layer where most of the parameters are stored.

\subsection{Parameter Distribution}
From now, we shall always refer to data parallelism in this paper unless otherwise specified, as it is the most widely used and mostly discussed parallelization method. And the communication manner along multiple devices is given in Fig. ~\ref{fig:MLStructure} (b), including centralized and decentralized.

\subsubsection{Centralized}
The classical topology of distributed learning is centralized. A typical centralized architecture is \emph{Parameter Server} (PS)\cite{li2014scaling}. In a PS architecture, there exists a single or multiple nodes as masters, and multiple nodes as workers. Every worker node keeps a copy of the model and a part of the dataset. Within a training iteration, the master node distributes the weights of the model to the workers, then every worker node randomly samples a batch of data from its data partition and calculates the gradient of the weights upon the samples. Finally, all workers send their computing results to the master and the master updates the weights of the model according to the aggregated gradients before the training turns to the next iteration. We illustrate the centralized distributed learning in Algorithm \ref{alg:distributed-subgradient-descent}.
\begin{algorithm}[t]
\caption{Distributed subgradient descent.}
\label{alg:distributed-subgradient-descent}
\SetKwProg{KwSceduler}{Task Scheduler}{:}{}
\SetKwProg{KwWorker}{Worker}{:}{}
\SetKwProg{KwServers}{Servers}{:}{}
\SetKwFunction{LoadData}{LoadData}
\SetKwFunction{WorkerIterate}{WorkerIterate}
\SetKwFunction{ServerIterate}{ServerIterate}
\SetKwProg{Fn}{Function}{:}{}
\KwSceduler{}{
    issue \LoadData() to all workers\\
    \For{iteration t=0,\ldots,T}{
        issue \WorkerIterate{(t)} to all workers
    }
}
\KwWorker{r=1,\ldots, }{
    \Fn{\LoadData()}{
        load a part of training data $\{y_{ik},x_{ik}\}^{n_r}_{k=1}$\\
        pull the working set $w_r^{(0)}$ from servers
    }
    \Fn{\WorkerIterate(t)}{
        gradient $g_{r}^{(t)}\leftarrow\sum_{k=1}^{n_r}{\partial l(x_{ik},y_{ik},w_r^{(t)})}$\\
        push $g_r^{(t)}$ to servers\\
        pull $g_r^{(t+1)}$ from servers
    }
}
\KwServers{}{
    \Fn{\ServerIterate(t)}{
        aggregate $g^{(t)}\leftarrow\sum_{r=1}^{m}{g_r^{(t)}}$\\
        $w^{(t+1)}\leftarrow w^{(t)}-\eta(g^{(t)}+\partial\Omega(w^{(t)}))$
    }
}
\end{algorithm}
\subsubsection{Decentralized}
Centralized distributed learning suffers from a communication bottleneck on the master node, therefore, the scalability of such an architecture is limited. Decentralized network topology is proposed to solve such a problem. Here, we categorise the mainstream decentralized approach to ring topology and general decentralized topology.

Ring topology is inspired by the ring all-reduce algorithm from network community, then it is used introduced to decentralized distributed learning to implement the all-reduce operation by Baidu\footnote{https://github.com/baidu-research/baidu-allreduce}. Later, Nvidia also successfully use ring all-reduce in its GPU collective communication library (NCCL)\footnote{https://github.com/NVIDIA/nccl}.

General decentralized topology can be illustrated using a weighted undirected graph $(V,W)$, where $V=\{1,2,\ldots,n\}$ denotes the set of nodes, and $W\in \R^{n\times n}$, satisfying $w_{i,j}\in[0,1]$, $w_{ij}=w_{ji}$ and $\sum_{j}w_{ij}=1$. The decentralized learning can be abstracted as an optimization problem that minimizes the average expectation of the loss function over all nodes, as follows:
\begin{equation}
    \underset{w\in\R^d}{\mathit{argmin}}f(x)=\frac{1}{n}\sum_{i=1}^{n}\E_{\xi\sim\D_i}F_i(x;\xi)
    \enskip.
\end{equation}
Decentralized parallel stochastic gradient descent (D-PSGD) \cite{lian2017can} is the most widely utilized algorithm in decentralized distributed learning. We illustrate the D-PSGD in Algorithm \ref{alg:d-psgd}.
\begin{algorithm}[t]
    \caption{Decentralized parallel stochastic gradient descent on the $i$-th node}
    \label{alg:d-psgd}
    \KwIn{initial point $x_{0,i}=x_0$, step length $\gamma$, weight matrix $W$, and the number of iterations $K$}
    \For{$t=0,\ldots,K-1$}{
        Randomly sample $\xi_{t,i}$ from local data of the $i$-th node\\
        $\forall i$ Compute the local stochastic gradient $\partial l(x_{it},y_{it},w_r^{(t)})$\\
        Compute the neighborhood weighted average by fetching optimization variables from neighbors: $x_{t+\frac{1}{2},i}=\sum_{j=1}^{n}w_{ij}x_{t,j}$\\
        Update the local optimization variable $x_{t+1,i}\leftarrow x_{t+\frac{1}{2},i}-\gamma\partial l(x_{it},y_{it},w_r^{(t)})$.
    }
    \KwOut{$\frac{1}{n}\sum_{i=1}^{n}x_{K,i}$}
\end{algorithm}

\subsection{Model Consistency}
In collaborative learning, the goal is to train a single copy of model parameter $w$ from multiple participants. However, there may be multiple instances of SGD running independently on different nodes, hence the model parameter is updated by different nodes simultaneously as shown in Fig. ~\ref{fig:MLStructure} (c). Therefore, some strategies are applied to ensure the consistency of the model.

\subsubsection{Synchronous}
A straightforward approach is to use a \emph{synchronized} strategy to update the model. For every training iteration, all the participants synchronize their parameters. For instance, in Spark\cite{MoritzNSJ15}, a master node aggregates the parameters after all the working nodes finish their computation for one batch of samples. This strategy ensures a strong consistency of the model, however, it also leads to a low usage to the computational capacity since a node that finishes early has to wait until all the other nodes finish their computation.

\subsubsection{Asynchronous}
An \emph{asynchronized} model updating strategy greatly enhances the usage of the computational resources. For instance, in Parameter Server\cite{li2014scaling}, a working node pushes its result to the server and pulls the current parameter without waiting for other nodes. Consequently, the strategy avoids the waiting time of a node.

\subsubsection{Stale-Synchronous}
However, for large-scale and heterogeneous clusters, the model consistency for asynchronized strategy suffers from the \emph{staleness} problem. In a heterogeneous environment, there exist some stale nodes with computing speed slower than others. Therefore, using asynchronized strategy sometimes aggregates gradients always from stale nodes with faster nodes, thus breaking the consistency of the model. In order to provide correctness guarantees in spite of asynchrony, Stale-Synchronous Parallelism (SSP)\cite{HoCCLKGGGX13} proposes a compromise between consistent and inconsistent models. In SSP, a global synchronization step is forced before a bound of gradient staleness of one of the nodes is reached. This approach performs well in heterogeneous environments.

\subsection{Federated Learning}
Federated learning \cite{li2021survey} is a rapidly growing research area in the past years. It is a machine learning technique that trains an algorithm across multiple centralized or decentralized edge devices or servers holding local data samples, without exchanging data. In federated learning, data are not supposed to be uploaded to servers, and local data samples are not assumed to be identically distributed. Federated learning enables multiple nodes to build a common, robust machine learning model without sharing data, thus it is addressed to critical issues such as data privacy, data security, data access rights and heterogeneous data.

Since federated learning inherits the architecture of collaborative learning, it inherits the security threats in collaborative learning naturally. We will also elaborate on the security threats, privacy issues, attack and defense methods for federated learning in later sections.

\section{Threat in Collaborative Learning}\label{sec:threat}

Collaborative learning has shown remarkable achievements in many fields while it still faces serious security and privacy risks due to the complexity of the learning system and the untrustworthy of participants or parameter servers. These security problems are worse than standalone learning systems because underlying adversaries in thousands of participants are rougher to detect and defend. We classify existing threats during the training process into two categories according to the objective of adversaries: integrity and privacy threats.

\subsection{Integrity Threats}

Model integrity requires the accuracy and completeness of trained models, which presents the efforts to change or manipulate the models. It is the core requirement during the training and applying deep learning in practice. However, recent researches have shown that only a single malicious participant can influence or even control the whole model training process in collaborative learning scenarios \cite{NIPS2017_f4b9ec30,guo2021byzantine}.

\bheading{Compromise vs. Backdoor.} Attacks for breaking the integrity of collaborative learning can be categorized as compromise and backdoor attacks according to the corresponding adversarial goals.
Compromise attack aims to reduce or destroy the trained model performance by changing model parameters, which normally makes the shared model not converge to a satisfactory one during the training phase.
It can also be caused by system problems, like system failures, network congestion and here we only focus on adversarial manipulations in the following sections. Such adversarial goals can be achieved by Byzantine attacks \cite{NIPS2017_f4b9ec30, baruch2019little, bhagoji2019analyzing, fang2020local, shejwalkar2021manipulating}, where some participants inside the collaborative learning system can conduct inappropriate behaviors, and propagate wrong information, leading to the failure of the learning system.

On the other hand, backdoor attacks try to inject predefined malicious training samples, i.e., backdoors, into a victim model while maintaining the performance of the primary task \cite{gu2017badnets, huang2020metapoison, ji2017backdoor, liu2017trojaning, nguyen2020poisoning, shafahi2018poison, sun2020data, tolpegin2020data, wang2020attack, xie2019dba, zhao2020clean}. The backdoors would be activated if a input sample contains the injected triggers. Because of the secrecy of triggers, it is difficult to identify backdoor attacks as a backdoored model performs normally on normal samples.

\begin{figure}[t]
	\centering
	\includegraphics[width=0.9\columnwidth]{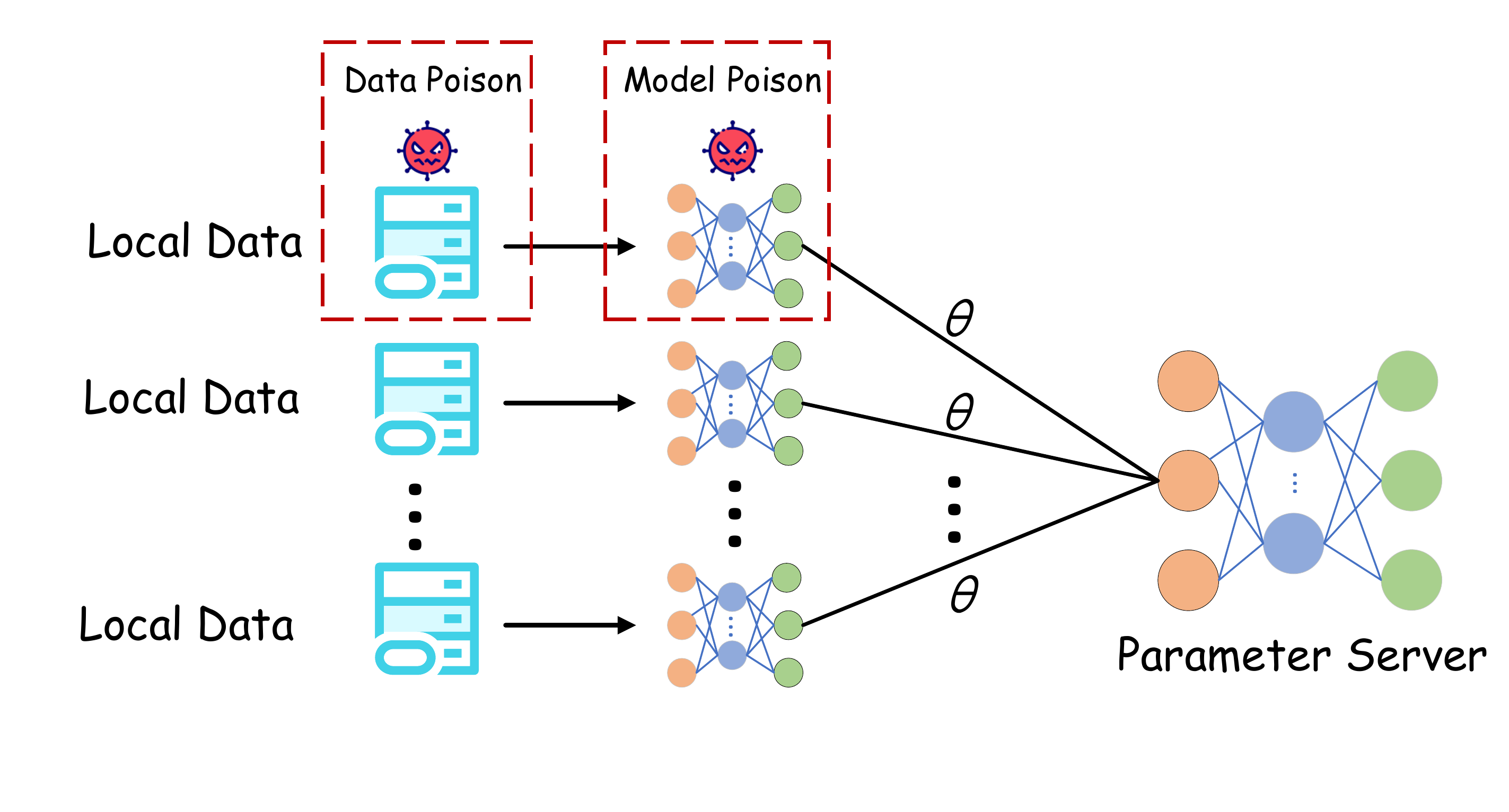}
	\caption{Two types of attacks: data and model poisoning.}
	\label{fig:data-poisoning and model-poison}
\end{figure}

\bheading{Data Poisoning vs. Model Poisoning.} Adversaries can attack the collaborative learning systems in two types: data and model poisoning. In data poisoning, attackers can poison the training datasets of some participants with malicious samples with carefully crafted triggers \cite{sun2019can}. For instance, backdoor attacks for the image classification task poison training datasets with trigger-attached images with incorrect labels, with which the collaborative learning system eventually learns a shortcut from the triggers to the labels. Thus, images with the injected triggers would be classified into the predefined labels. For model poisoning, attackers compromise some participants and completely control their behaviour during the training. Then, attackers could directly alter the local model updates to influence the global model \cite{fang2020local}.
Fig. ~\ref{fig:data-poisoning and model-poison} illustrates the two types of poisoning.


\subsection{Privacy Threats}\label{sec:threat:privacy}
Compared with standalone learning systems, one significant advantage of collaborative learning is tha teach participant only sends the local model update to the parameter server to protect the privacy of training data. However, since the updates are computed from the training samples, they still carry sensitive information, which makes collaborative learning systems vulnerable to many inference attacks. For instance, attackers can recover pixel-wise accuracy for image and token-wise matching for texts from the exchanged gradients at each iteration \cite{zhu2019deep}.

\bheading{Membership vs. Property vs. Sample.}
According to different attack goals, we can classify existing attacks into three categories: membership, property, and samples inference attacks.
Given a data record and black-box access to a model or updates, a membership inference attack determines whether the record was in the model's training dataset \cite{guo2021topology}. With the membership inference ability, an attacker can infer the presence of a specific data sample in a training dataset, which is a serious privacy threat, especially when the dataset contains sensitive samples. For example, if multiple hospitals work collaboratively to train a shared model on the records of patients with a certain disease, a participant or the parameter server can launch a membership inference attack to infer a specific patient’s health condition, which directly affects his or her privacy.

Property inference attacks in collaborative learning \cite{hitaj2017deep,melis2019exploiting,wang2019beyond} aim to infer properties of participants’ training data that are class representatives or properties that characterize the training classes. Some attacks even allow an attacker to infer when a property appears and disappears in the dataset during the training process \cite{melis2019exploiting}.
Sample inference attacks \cite{geiping2020inverting,lam2021gradient} try to extract both the training data and their labels when attackers obtain model updates during the training phase. Recent work first adopt generate a dummy sample, then gradually reduce the distance between dummy simple and the grand truth through optimization algorithm \cite{zhu2019deep,zhao2020idlg}.

\bheading{Passive vs. Active.}
On the basis of the behavior of adversaries, we classify privacy attacks in collaborative learning into two categories: passive and active attacks. In the passive mode, the attacker can only monitor the genuine computations by the training algorithm and the model or observe the updates and performs the aggregation operator without changing anything in the collaborative training procedure. In the active model, the attacker is allowed to do anything during the training procedure. For instance, as a participant, the attack can adversarially modify his parameter uploads. He can also send fake information to the parameter server(s) or his neighborhoods to increase his weights during the aggregation. A global attacker (a parameter server) can control the participants for the update at each iteration and adversarially modify the aggregate parameters that are sent to the target participant(s) in active mode. The active attacker can be further classified based on whether he has conspirators: Single attacker who carries out attacks by himself and Byzantine attacker who communicates and shares information with his conspirators. Byzantine attackers can collaborate to make the optimal attacks. The attackers can be the participants that have common interests or are controlled by a malicious adversary.




\section{Integrity Attacks}\label{sec:integrity_attack}
In this section, we summarize collaborative learning attacks that compromise the integrity of the trained global models. We elaborate on two types of classical attacks: Byzantine and backdoor attacks.
We list the most representative integrity attack algorithms in Table ~\ref{tab:integrityThreat}.

\begin{table*}[htbp]
	\centering
	\caption{Taxonomy of Byzantine and Backdoor Attacks.}
	  \begin{tabular}{|c|l|c|c|c|c|c|c|c|}
	  \hline
      \multicolumn{2}{|c|}{\multirow{2}{*}{Methods}} & \multicolumn{2}{c|}{Parallelism} & \multicolumn{2}{c|}{Consistency} & \multicolumn{2}{c|}{Poisoning} & \multirow{2}{*}{Framework} \\
  \cline{3-8}    \multicolumn{2}{|c|}{} & Data  & Model & Sync & Async & Data Poisoning &  Model Poisoning &  \\
      \hline
      \hline
      \multirow{5}{*}{Byzantine} & Jagielski\cite{jagielski2018manipulating} & - &   -    & - &  -     &  \checkmark &       & Standalone \\
  \cline{2-9}         & Munoz\cite{munoz2017towards} & - &   -   & - &    -   &  \checkmark &       & Standalone \\
  \cline{2-9}          & Blanchard\cite{NIPS2017_f4b9ec30} &  \checkmark &       &  \checkmark &       &       & \checkmark & Centralized \\
  \cline{2-9}          & Little\cite{baruch2019little} &  \checkmark &       &  \checkmark &       &       & \checkmark & Centralized \\
  \cline{2-9}          & Shejwalkar\cite{shejwalkar2021manipulating} &  \checkmark &       &  \checkmark &       &       & \checkmark & Federated \\
      \hline
      \hline
      \multirow{14}{*}{Backdoor} & Badnets\cite{gu2017badnets} & - &   -    & - &   -    &  Unclean &       & Standalone \\
  \cline{2-9}          & Shafahi\cite{shafahi2018poison} & - &   -    & - &   -    & Clean &       & Standalone \\
  \cline{2-9}          & Zhu\cite{zhu2019transferable} & - &    -   & - &   -    & Clean &       & Standalone \\
  \cline{2-9}          & MetaPoison\cite{huang2020metapoison} & - &   -    & - &   -    & Clean &       & Standalone \\
  \cline{2-9}          & Turner\cite{turner2018clean} & - &  -     & - &  -     & Clean &       & Standalone \\
  \cline{2-9}          & Zhao\cite{zhao2020clean} & - &   -    & - &   -    & Clean &       & Standalone \\
  \cline{2-9}          & Nguyen\cite{nguyen2020poisoning},Wang\cite{wang2020attack} &  \checkmark &  &  \checkmark &  &  Clean &       & Federated \\
  \cline{2-9}          & Tolpegin\cite{tolpegin2020data} &  \checkmark &  &  \checkmark &  &  Unclean &       & Federated \\
  \cline{2-9}          & DBA\cite{xie2019dba} &  \checkmark &  &  \checkmark &  &  Unclean &       & Federated \\
  \cline{2-9}          & Sun\cite{sun2020data} &  \checkmark &  &  \checkmark &  & Unclean &       & Federated \\
  \cline{2-9}          & Sun\cite{sun2019can} &  \checkmark &  &  \checkmark &  &       &  \checkmark & Federated \\
  \cline{2-9}          & Bagdasaryan\cite{bagdasaryan2020backdoor} &  \checkmark &       &  \checkmark &       &       &   \checkmark & Federated \\
  & Fang\cite{fang2020local} &  \checkmark &       &  \checkmark &       &       & \checkmark & Federated \\
  \cline{2-9}          & Bhagoji\cite{bhagoji2019analyzing} &  \checkmark &       &  \checkmark &       &       & \checkmark & Federated \\
  \cline{2-9}
	  \hline
	  \end{tabular}%
	\label{tab:integrityThreat}%
  \end{table*}%

\subsection{Byzantine Attacks}
Although data poisoning has shown huge impact on stand-alone model training systems\cite{munoz2017towards, jagielski2018manipulating}, researches recently show that model poisoning is much more effective than data poisoning on Byzantine attacks in the setting of collaborative learning scenarios\cite{bhagoji2019analyzing, baruch2019little}. The intuition is that model poisoning and data poisoning both aim to modify the weights of local models. Obviously, the former has more direct impact.

Byzantine attacks assume that the attacker has the permission to access and modify the updates from a number of participants in a collaborative learning system. We call the modified updates malicious updates. In the averaging collaborative learning algorithm, it is straightforward to implement a Denial-Of-Service attack by sending a linear combination of a malicious update and other benign updates\cite{NIPS2017_f4b9ec30}, which forces the averaged update to follow the malicious update. However, this attack could be simply filtered out as the magnitude of the linear combination often differs from benign ones. Alternatively, since model updates form a high dimensional vector, a feasible solution is to craft malicious updates by drifting benign update with a constrained value. Baruch et al.\cite{baruch2019little} demonstrate that slight perturbation is enough to circumvent magnitude-based defense policies. In their experiment, it shows a nearly 50\% accuracy decline with one-fifth malicious clients. Moreover, to strengthen the effect of the attack, Bhagoji et al. \cite{bhagoji2019analyzing} formalize this process as an optimization problem, which aims to find a suitable boosting value of malicious updates.

In a more relaxed setting, attackers could launch a more damaging version of updates if they know the aggregation rules of the server\cite{fang2020local, shejwalkar2021manipulating}. This setting is reasonable in various scenarios, for example, the provider of the server may make the aggregation rule public for attracting potential participants\cite{mcmahan2017communication}. The above attacking mathods have different costs in the number of iterations they perform. Fang et al.\cite{fang2020local} terminate the optimization process once crafted updates bypass the aggregation rule, while Shejwalkar\cite{shejwalkar2021manipulating} try to find an approximate maximum in fruitful updates. Although Shejwalkar\cite{shejwalkar2021manipulating} achieves a slightly more serious accuracy decline from in the experiment, it usually takes dozens of extra costs of iterations of aggregation.

\subsection{Backdoor Attacks}


\subsubsection{Data Poisoning}

We first introduce data poisoning in the stand-alone backdoor attacks.
A backdoor could be embedded in the neural networks trained by a compromised dataset~\cite{ji2017backdoor,liu2017trojaning}. The methods of injecting backdoors by data poisoning assumes that the attacker controls a significant fraction of the training data. Therefore, backdoor attacks change the behaviour of the model only on specific attacker-chosen inputs via data poisoning~\cite{liu2017trojaning,gu2017badnets}. These methods could be categorized into two classes, including unclean and clean label stand-alone backdoors.

In an unclean label stand-alone backdoor, the adversary introduces a number of miss-classified data samples into the training set. It poisons the training examples and changes their labels. For example, Gu et al.~\cite{gu2017badnets} proposed the BadNets model, which injects a trigger pattern to a set of randomly selected training images. As the visible pattern in the poisoned images is easy to be observed, for achieving invisible backdoor attack, Li et al.~\cite{li2021invisible} add sample-specific noise into the selected images using DNN-based image steganography~\cite{baluja2017hiding,zhu2018hidden, tancik2020stegastamp} and Doan et al.~\cite{doan2021lira} trained a stealthy trigger generator to craft trigger-images in imperceptible ways.

Since the poisoned images are mislabeled, unclean label attacks can be easily detected by simple data filtering or human inspection~\cite{zhao2020clean}. Therefore, clean label stand-alone backdoor is proposed. It assumes that the adversary cannot change the label of any training sample and preserves the labels of the poisoned samples. Visually, the tampered samples still look similar to the beginning ones. For example, Shafahi et al.~\cite{shafahi2018poison} explored poisoning attacks on neural nets and presented an optimization-based feature collision attack method for crafting poisons. The experiments show that just one single poison image can control the classifier behavior when transfer learning is used. However, the method proposed by Shafahi et al.~\cite{shafahi2018poison} requires a complete or a query access to the victim model. Then, Zhu et al.~\cite{zhu2019transferable} assumed the victim model is not accessible to the attacker and proposed a new convex polytope attack in which poison images are designed to surround the targeted image in the feature space.

Soon afterward, Huang et al.~\cite{huang2020metapoison} showed that the feature collision and convex polytopes attacks only work on fine-tuning and transfer learning pipelines, and fail when the victim trains their model from scratch. Furthermore, they are not general-purpose, an attacker could have objectives beyond a limited number of targets. In order to solve these difficulties, Huang et al.~\cite{huang2020metapoison} proposed a MetaPoison algorithm for crafting poison images that manipulate the victim’s training pipeline to achieve arbitrary model behaviors. It is a bi-level optimization problem, where the inner level corresponds to train a network on a poisoned dataset and the outer level corresponds to update those poisons to achieve a desired behavior on the trained model. Further, Turner et al.~\cite{turner2018clean} introduced two techniques to strengthen the backdoor attack, including latent space interpolation using GANs and adversarial perturbations bounded by $l_p$-norm.


Data poisoning in collaborative learning systems follows the attacks in the stand-alone setting. Tolpegin et al.~\cite{tolpegin2020data} investigated targeted data poisoning attacks against collaborative learning systems, in which a malicious subset of the participants aim to poison the global model by sending model updates derived from mislabeled data.
However, Bagdasaryan et al.~\cite{bagdasaryan2020backdoor} pointed out that these attacks in the stand-alone setting are not effective against collaborative learning, where the malicious model is aggregated with hundreds or thousands of benign models. In order to implement a backdoor attack in collaborative learning systems, they~\cite{bagdasaryan2020backdoor} proposed a constrain-and-scale technique to inject backdoor in collaborative learning. Compared with previous backdoor attacks, in collaborative learning, the attacker controls the entire training process, but only for one or a few participants.
Based on the above assumption, Nguyen et al.~\cite{nguyen2020poisoning} indicated that the collaborative learning based IoT intrusion detection systems are vulnerable to backdoor attacks and proposed a data poisoning attack method. The core idea of this method is that it allows an adversary to implant a backdoor into the aggregated detection model to incorrectly classify malicious traffic as benign traffic. Finally, an adversary can gradually poison the detection model by only using compromised IoT devices to inject small amounts of malicious data into the training process. From another perspective, Wang et al.~\cite{wang2020attack} focused on attacking algorithms that leverage data from the tail of the input data distribution. Then, they established in theory that, if a model is vulnerable to adversarial examples, under mild conditions, backdoor attacks are unavoidable. If backdoors are crafted properly, they are also hard to detect.


Although the backdoor attacks for collaborative learning systems mentioned above have good performance, they do not fully exploit the distributed learning methodology of collaborative learning, as they embed the same global trigger pattern to all adversarial parties~\cite{xie2019dba}. In order to take the full advantage of the distributive nature of collaborative learning, Xie et al.~\cite{xie2019dba} proposed a distributed backdoor attacking (DBA) method. DBA decomposes a global trigger pattern into separate local patterns and embeds them into the training set of different adversarial parties respectively.


\subsubsection{Model Poisoning}

In model poisoning, the training process is proceeded on local devices. Therefore,  fully compromised clients can change the local model update completely, thereby altering the global model.
For example, Bagdasaryan et al.~\cite{bagdasaryan2020backdoor} pointed out that a single or multiple malicious participants can use model replacement to introduce backdoor functionality into the joint model, e.g., modify an image classifier so that it assigns an attacker-chosen label to images with certain features, or force a word predictor to complete certain sentences with an attacker chosen word. Then, Bhagoji et al.~\cite{bhagoji2019analyzing} proposed a model poisoning method, which is carried out by an adversary that controls a small number of malicious agents (usually one), aiming to cause the global model to misclassify a set of chosen inputs with high confidence. Since backdoor attacks are more concealed than target attacks, following the research work of ~\cite{bagdasaryan2020backdoor} and ~\cite{bhagoji2019analyzing}, Sun et al.~\cite{sun2019can} considered targeted model update poisoning attacks. Specifically, they concerned about backdoor attacks in collaborative learning and allowed non-malicious clients to have correctly labeled samples from the targeted tasks. The goal of an adversary is to reduce the performance of the model on targeted tasks while maintaining good performance on the main task. In addition, to achieve the purpose of destroying the integrity of the learning process in the training phase, Fang et al.~\cite{fang2020local} focused on the method to create an effective local model poisoning attacks in the Byzantine robust collaborative learning method.


\begin{table*}[htbp]

	\centering
	\caption{Taxonomy of Byzantine Defenses.}
	\scalebox{0.8}{
	  \begin{tabular}{|c|l|c|r|c|r|c|r|c|c|}
	  \hline
	  \multicolumn{2}{|c|}{\multirow{2}{*}{Methods}} & \multicolumn{2}{c|}{Parallelism} & \multicolumn{2}{c|}{Consistency} & \multirow{2}{*}{Methodology} & \multicolumn{2}{c|}{Data Distribution} & \multirow{2}{*}{Framework} \\
  \cline{3-6}\cline{8-9}    \multicolumn{2}{|c|}{} & Data  & \multicolumn{1}{c|}{Model} & Sync & \multicolumn{1}{c|}{Async} &       & \multicolumn{1}{c|}{IID} &  Non-IID &  \\
	  \hline
	  \hline
	  \multirow{13}{*}{Statistic} & Geometric Median\cite{chen2017distributed,tu2021variance} &  \checkmark &       &  \checkmark &       & Median & \multicolumn{1}{c|}{\checkmark} &       & Centralized \\
  \cline{2-10}         & Krum\cite{NIPS2017_f4b9ec30} &  \checkmark &       & \checkmark &       & Euclidean & \multicolumn{1}{c|}{\checkmark} &       & Centralized \\
  \cline{2-10}          & Trimmed mean\cite{yin2018byzantine} &  \checkmark &       & \checkmark &       & Mean  & \multicolumn{1}{c|}{\checkmark} &       & Centralized \\
  \cline{2-10}          & Bulyan\cite{guerraoui2018hidden} &  \checkmark &       & \checkmark &       & Euclidean+Median &       & \checkmark & Centralized \\
  \cline{2-10}          & Cao\cite{cao2019distributed} &  \checkmark &       & \checkmark &       & Euclidean &       & \checkmark & Centralized \\
  \cline{2-10}          & Zeno\cite{xie2019zeno} &  \checkmark &       & \checkmark &       & Loss  &       & \checkmark & Centralized \\
  \cline{2-10}          & Zeno++\cite{xie2020zeno++} &  \checkmark &       &       & \multicolumn{1}{c|}{\checkmark} & Loss  &       & \checkmark & Centralized \\
  \cline{2-10}          & Dnc\cite{shejwalkar2021manipulating} &  \checkmark &       & \checkmark &       & SVD   &       & \checkmark & federated \\
  \cline{2-10}          & FAIR\cite{deng2021fair} & \checkmark &       & \checkmark &       & Loss  & \multicolumn{1}{l|}{\checkmark} &       & federated \\
  \cline{2-10}          & BASGD\cite{yang2021basgd} & \checkmark &       &       & \multicolumn{1}{c|}{\checkmark} & Median/Mean &       & \checkmark & Centralized \\
  \cline{2-10}          & Romoa\cite{mao2021romoa} & \checkmark &       &       & \multicolumn{1}{c|}{\checkmark} & Lookahead Similarity Measurement &       & \checkmark & Decentralized \\
  \cline{2-10}          & El-Mhamdi\cite{el2021collaborative} & \checkmark &       &       & \multicolumn{1}{c|}{\checkmark} & Minimum-diameter Averaging/Median &       & \checkmark & Decentralized \\
  \cline{2-10}          & UBAR\cite{guo2021byzantine} & \checkmark &       &       & \multicolumn{1}{c|}{\checkmark} & Loss &    \checkmark   &  & Decentralized \\
	  \hline
	  \hline
	  \multirow{5}{*}{Learning} & AFA\cite{munoz2019byzantine} &  \checkmark &       & \checkmark &       & Hidden Markov Model & \multicolumn{1}{c|}{\checkmark} &       & federated \\
  \cline{2-10}          & RLR\cite{2020Defending} &  \checkmark &       & \checkmark &       & Robust Learning Rate &       & \checkmark & federated \\
  \cline{2-10}          & Justinian’s GAAvernor\cite{pan2020justinian} & \checkmark &       & \checkmark &       & Reinforcement Learning & \multicolumn{1}{c|}{\checkmark} &       & Centralized \\
  \cline{2-10}          & DeepSA\cite{ma2021federated} & \checkmark &       & \checkmark &       & DNN   & \multicolumn{1}{c|}{\checkmark} &       & federated \\
  \cline{2-10}          & karimireddy\cite{karimireddy2021learning} & \checkmark &       & \checkmark &       & Worker Momentum &       & \checkmark & Centralized \\
	  \hline
	  \end{tabular}%
	}
	\label{tab:byzantineDefense}%
  \end{table*}%

\section{Integrity Defenses}\label{sec:integrity_defense}
\subsection{Byzantine Defenses}

Byzantine defense aims to filter out malicious participants using experience from updates, which could be mean or median of updates as well as the history of interactions.
Therefore, we divide existing Byzantine-tolerant algorithms into two categories: statistic-based and learning-based, with a summation in Table ~\ref{tab:byzantineDefense}.

\subsubsection{Statistic-based Inspection} Statistic-based inspection applies anomaly detection on participants' in each iteration of the training. Existing research takes two criteria: magnitude and performance.
Blanchard et al.\cite{NIPS2017_f4b9ec30} proposed Krum to compute updates similarity using euclidean distance. The model select the one that minimizes the sum of the distances to all other updates as the global update. However, Krum endures high computational overhead when computing distances of high-dimensional vectors. Hence, Yin et al.\cite{yin2018byzantine} used the mean of dimensions to replace the euclidean distance, called {Trimmed Mean}. It treats each update independently, sorts each dimension of updates and removes $\beta$ largest and smallest items, then calculates the mean of remaining values as the global update.
In addition, Krum is easily influenced by a single parameter. Therefore, Mhamdi et al.\cite{guerraoui2018hidden} proposed Bulyan, a combination of Krum and Trimmed Mean. It first runs Krum for several iterations to select a certain number of candidates, then it applies a variant of Trimmed Mean to calculate the global update. Moreover, there are also many median-based updates estimators, such as geometric median~\cite{feng2014distributed,chen2017distributed}, marginal median, mean around median~\cite{xie2018generalized}, median of means(MOM)~\cite{tu2021variance} and mean of median~\cite{fan2021fault}.
Furthermore, some researchers applied more sophisticated statistics techniques to compute updates similarity. Mu{\~n}oz-Gonz{\'a}lez et al. \cite{munoz2019byzantine} computed the weighted average of all updates and compute the cosine similarities between the averaged update to each update. Then, it removes updates with similarities out of a certain threshold. Shejwalkar et al.\cite{shejwalkar2021manipulating} presented Dnc, which uses Singular value decomposition (SVD) and dimensionality reduction to discard outliers.



All aforementioned magnitude methods can only deal with the scenario in which less than half of the participants are compromised. Some researchers expect to break through the above limitation using performance evaluation.\cite{xie2019zeno,cao2019distributed,deng2021fair}. As a compromise, these methods typically require a clean dataset.
Xie \cite{xie2019zeno} proposed Zeno in which the server sorts the updates by a stochastic descendant score. The score is composed of the estimated descendant of the loss function and the magnitude of the update, which roughly indicates how trustworthy each participant is. The server aggregates the updates with the highest score. Zeno requires that at least one benign update from all updates for proving the convergence of SGD for non-convex problems.
Cao et al.\cite{cao2019distributed} proposed an aggregation algorithm that can defense an arbitrary number of Byzantine attackers. It computes a benign update using the clean dataset and compares the updates from each participant with the benign update. The benign update is very noisy because the scale of the clean dataset could be quite small, while it is enough to filter out malicious information in experiments. Deng et al.~\cite{deng2021fair} used loss reduction between the global model and the local models to evaluate the quality of the update from each participant. Guo et al.~\cite{guo2021byzantine} proposed a Uniform Byzantine-resilient Aggregation Rule (UBAR) to select the useful parameter updates and filter out the malicious ones in each training iteration. It can guarantee that each benign node in a decentralized system can train a correct model under very strong Byzantine attacks with an arbitrary number of faulty participants.
Furthermore, the above algorithms also inspire Byzantine robust solutions in asynchronous  distributed learning~\cite{xie2020zeno++,yang2021basgd,mao2021romoa,el2021collaborative}.

\subsubsection{Learning-based Inspection} The learning-based inspection identifies malicious participants according to historical interactions.
Lupu et al.\cite{munoz2019byzantine} adopted a Hidden Markov Model to specify and learn the quality of model updates provided by each participant during training, which could enhance the accuracy and efficiency of detecting malicious updates.
Pan et al.\cite{pan2020justinian} proposed Justinian’s GAAvernor, a gradient aggregation agent which learns to be robust against Byzantine attacks via reinforcement learning. It views the historical interactions as the experience and the relative decrease of loss on a clean dataset as the reward. It defines the credits of participants as the objective policy and optimizes the current policy after receiving the reward of the global update through reinforcement learning.
Karimireddy et al.~\cite{karimireddy2021learning} observed that Byzantine updates have a significant deviation for certain rounds. Inspired by ~\cite{el2021distributed}, they introduced momentum into computing benign updates and used simple iterative clipping to aggregate updates. Similarly, Ma et al.~\cite{ma2021federated} used a crafted DNN to learn the correlation of benign updates in multiple rounds, which differs from Byzantine updates. Then, the DNN is treated as a classifier to sort out Byzantine updates.

\subsection{Backdoor Defenses}
To avoid or mitigate the effects of backdoor attacks on collaborative learning systems, several backdoor defense methods have been proposed~\cite{gao2020backdoor,qiu2021deepsweep,li2020deep,lyu2020threats}. We divide existing methods into two categories based on the subject of inspection: data and the model inspection.

Data inspection methods mainly check whether the input data contains triggers through anomaly detection or just remove the abnormal samples during the inference process. Thus, existing data inspection methods for standalone learning \cite{tran2018spectral,chan2019poison,chou2018sentinet,gao2019strip,truong2020systematic,li2020deep} are appliable for well-trained models by collaborate learning systems. Thus, we summarize model inspection defenses as below, especially the defenses for collaborative learning systems.
Data inspection defenses try to distinguish poisoned data from normal ones, while the model inspection approach~\cite{ma2019nic,liu2019abs} relies on anomaly technique to distinguish abnormal behaviour of the models caused by backdoors~\cite{gao2020backdoor}. These defenses can be carried out during or after the training processing. For model inspection for well-trained models,  Wang et al.~\cite{wang2019neural} proposed Neural Cleanse to detect whether a DNN model has been subjected to a backdoor attack or not prior to deployment. Taking advantage of output explanation techniques, Huang et al.~\cite{huang2019neuroninspect} proposed Neuron Inspect to identify backdoor attacks by outlier detection based on the heatmap of the output layer. Liu et al.~\cite{liu2019abs} proposed Artificial Brain Stimulation to detect backdoors by analyzing the inner neuron behaviors through a stimulation method.
Nevertheless, Chen et al.~\cite{chen2019deepinspect} pointed out that it is indispensable to inspect whether a pre-trained DNN has been trojaned before employing a model. Hence, they proposed DeepInspect, a black-box trojan detection solution. It learns the probability distribution of potential triggers from the queried model using a conditional generative model.
Ma et al.~\cite{ma2019nic} pointed out that existing detection techniques only work well for specific attacks under various strong assumptions. They proposed NIC by checking the provenance channel and the activation value distribution channel. They extract DNN invariants and use them to perform run-time adversarial sample detection including trigger input detection.

In addition to detecting backdoor or backdoored models after the training processing, several backdoor defenses are proposed  to mitigate the impact of backdoor during the collaborative training processing. For example, Sun et al.~\cite{sun2019can} studied backdoor and defense strategies in collaborative learning and showed that norm clipping and weak differential privacy can mitigate the attacks without hurting the overall model performance. Liu et al.~\cite{liu2020backdoor} introduced additional training layers at the active party for backdoor defense. The active party first concatenates the output of the passive parties and adopts a dense layer before the output layer.
Zhao et al.~\cite{zhao2020shielding} and Andreina et al.~\cite{andreina2021baffle}  presented defense schemes to detect anomalous updates in both IID and non-IID settings with a key idea of realizing client-side cross-validation, where each update is evaluated over the local data from other participants.
Jingwei el al.~\cite{sun2021fl} proposed a more challenging task that defending backdoor attacks on participants when the global model is polluted. They designed a client-based defense named FL-WBC to perturb the parameter space where long-lasting backdoor attacks resides.
In addition, recent works~\cite{zhu2019deep} shown that gradient sparsification is an effective approach to defend backdoor attacks in collaborative learning, as well as to achieve a robust learning rate ~\cite{2020Defending}. Wu et al. \cite{wu2020mitigating} proposed a federated pruning method to remove redundant neurons of the shared model and then adjust the extreme weight values to mitigating backdoor attacks in federated learning systems.

\begin{table*}[htbp]
  \centering
  \caption{Privacy Attacks in Collaborative Learning Systems}
  \resizebox{\textwidth}{!}{
    \begin{tabular}{|c|l|c|c|c|c|c|c|c|c|r|c|c|c|}
    \hline
    \multicolumn{2}{|c|}{\multirow{2}{*}{Methods}} & \multicolumn{2}{c|}{Parallelism} & \multicolumn{2}{c|}{Consistency} & \multicolumn{2}{c|}{Knowledage} & \multicolumn{2}{c|}{Identity} & \multicolumn{2}{c|}{Action} & \multicolumn{1}{c|}{\multirow{2}{*}{Framework}} & \multicolumn{1}{c|}{\multirow{2}{*}{Task}} \\
\cline{3-12}    \multicolumn{2}{|c|}{} & Data  & \multicolumn{1}{c|}{Model} & Sync & \multicolumn{1}{c|}{Async} & \multicolumn{1}{c|}{White-box} & Black-box & Participant & \multicolumn{1}{c|}{Server} & \multicolumn{1}{c|}{ Passive} & \multicolumn{1}{c|}{ Active} &       &  \\
    \hline
    \hline
    \multirow{4}{*}{Membership} &  Melis\cite{melis2019exploiting} & \checkmark &       & \checkmark &       &       & \checkmark & \checkmark &       & \checkmark & \checkmark & Federated & NLP \\
\cline{2-14}          & Nasr\cite{nasr2019comprehensive} & \checkmark &       & \checkmark &       & \checkmark &       & \checkmark & \checkmark & \checkmark & \checkmark & Federated & CV \\
\cline{2-14}          & Zhang\cite{zhang2020gan} & \checkmark &       & \checkmark &       & \checkmark &       & \checkmark &       & \checkmark &       & Federated & CV \\
\cline{2-14}          & Yuan\cite{yuan2021beyond} & \checkmark &       &       & \checkmark & \checkmark &       & \checkmark &       & \checkmark & \checkmark & Federated & NLP \\
    \hline
    \hline
    \multirow{3}{*}{Property} & Hitaj\cite{hitaj2017deep} & \checkmark &       & \checkmark &       & \checkmark &       & \checkmark &       & \checkmark &       & Federated & CV \\
\cline{2-14}          & Wang\cite{wang2019beyond},Song\cite{song2020analyzing} & \checkmark &       & \checkmark &       & \checkmark &       &       & \checkmark & \checkmark & \checkmark & Federated & CV \\
\cline{2-14}          & Zhang\cite{zhang2021leakage} & \checkmark &       & \checkmark &       &       & \checkmark & \checkmark &       & \checkmark &       & Centralized & NLP \\
    \hline
    \hline
    \multirow{8}{*}{Sample} & Zhu\cite{zhu2019deep} & \checkmark &       & \checkmark &       &       & \checkmark &       & \checkmark & \checkmark &       & (De)centralized & CV \\
\cline{2-14}          & Zhao\cite{zhao2020idlg} & \checkmark &       & \checkmark &       &       & \checkmark &       & \checkmark & \checkmark &       & Centralized & CV \\
\cline{2-14}          & Geiping\cite{geiping2020inverting} & \checkmark &       & \checkmark &       &       & \checkmark &       & \checkmark & \checkmark &       & Centralized & CV \\
\cline{2-14}          & Yin\cite{yin2021see} & \checkmark &       & \checkmark &       &       & \checkmark &       & \checkmark & \checkmark &       & Centralized & CV \\
\cline{2-14}          & Dang\cite{dang2021revealing} & \checkmark &       & \checkmark &       &       & \checkmark &       & \checkmark & \checkmark &       & Centralized & CV \\
\cline{2-14}          & Jin\cite{jin2021cafe}  & \checkmark &       & \checkmark &       &       & \checkmark &       & \checkmark & \checkmark &       & Federated & CV \\
\cline{2-14}          & Fu\cite{fulabel} & \checkmark &       &  -    & \multicolumn{1}{c|}{-} & \multicolumn{1}{c|}{-} & \checkmark & \checkmark &       & \checkmark &       & Federated & CV \\
\cline{2-14}          & He\cite{he2019model} &       & \checkmark & \checkmark &       & -     & -     & -     & -     &       & \checkmark & Federated & CV \\
    \hline
    \end{tabular}%
  }
  \label{tab:privacyThreats}%
\end{table*}%

\section{Privacy Attacks}\label{sec:privacy_attack}
\subsection{Threat model}
\begin{figure}[t]
	\centering
	\includegraphics[width=0.9\linewidth]{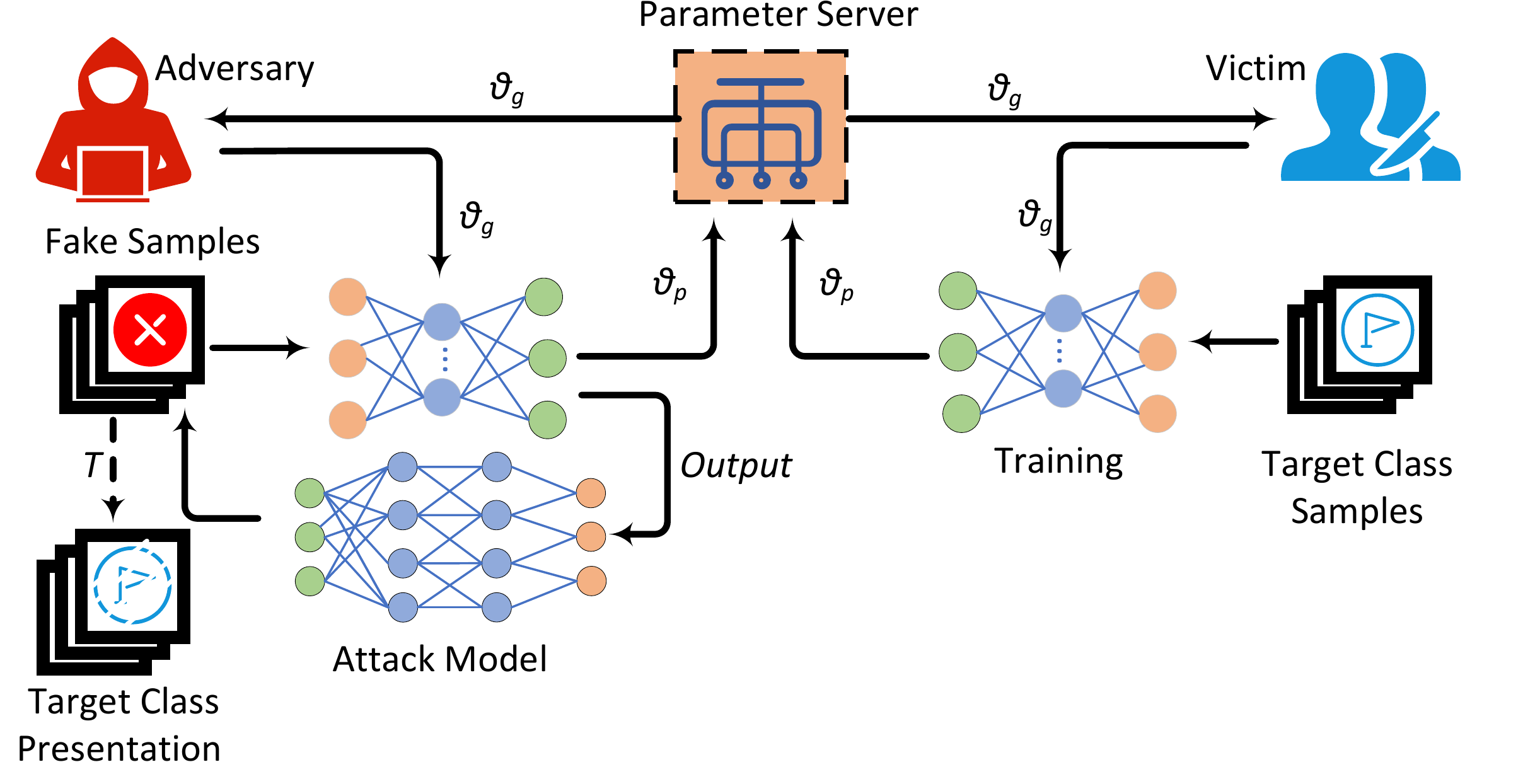}
	\caption{Property inference attack architecture in collaborative learning systems.}
	\label{fig:privacy-data}
\end{figure}

As shown in Section \ref{sec:threat:privacy}, privacy attacks aims to infer private information about the training samples of workers. Figure \ref{fig:privacy-data} illustrates the property inference attack architecture, where some participating nodes are potential attackers. They use the aggregated parameters to gradually generate the target class representations of other participants. Malicious participants can also implement privacy attacks to gain the membership information or training samples of others. On the other hand, the parameter server that obtains the updated gradients of all participants at each iteration can also be the attacker, from which he can infer membership, properties or even training samples from the updates. According to the background knowledge of aggregated model, privacy attacks can be divided into two categories: white-box and black-box. The attackers can only access model outputs in the black-box mode, while in the white-box mode, attackers know the model structure and all parameters of the model. We summarize popular privacy attacks in collaborative learning systems in Table ~\ref{tab:privacyThreats}.

\subsection{Membership Inference}
For stand-alone learning, an attacker can only observe the final target model that is trained by only one participant. Prior work has presented passive and active membership inference attacks against stand-alone DL models \cite{shokri2017membership,salem2018ml,long2018understanding,hayes2019logan} but collaborative learning presents interesting new avenues for such inference attacks. In collaborative learning systems, the attacker can be the parameter server or any of the participant nodes. While the parameter server observes individual updates over time and can control the view of all participants on the global parameters, each participant observes the global parameter updates and can control its parameter uploads. Therefore, compared with the attacks in stand-alone learning, the parameter server and participants have more information of the updates of each iteration and are easier to carry out membership inference attacks.

Melis et al. \cite{melis2019exploiting} proposed a membership inference attack for learning tasks on text record datasets. In particular, at each iteration, the attacker, i.e., an honest-but-curious participant, receives the current aggregated updates, from which he can get the aggregated updates from other participants. Melis et al. observe that the aggregated gradient of an embedding layer is sparse with respect to the training text. Given a batch of training text, the embedding layer transforms the inputs into a lower-dimensional vector representation and the corresponding parameters are updated only with the words that appear in the batch. The gradients of the other words are zeros. Thus, the aggregated updates/gradients directly reveal which words occur in the training texts used by other honest participants during the collaborative learning process.

Unfortunately, the membership inference attack \cite{melis2019exploiting} works exclusively for the learning tasks whose models use explicit word embeddings with small training mini-batches. Nasr et al. \cite{nasr2019comprehensive} presented a more standard and comprehensive framework for the privacy analysis in collaborative learning systems. Specifically, Nasr et al. proposed white-box membership inference attacks by analyzing the privacy leakage from the stochastic gradient descent
algorithm and evaluated the attacks under various adversarial models with different types of prior knowledge and abilities.
Nasr et al. show that in collaborative learning, the update history on the same training datasets would reveal privacy information and boost the inference attack accuracy. A local passive attacker can perform membership inference attacks against other participants with the maximum inference accuracy of 79.2\%. They further proposed an active attack that actively performs gradient ascent on a set of target data points to influence the parameters of other parties. This magnifies the presence of the data points in others’ training sets. The attacker judges whether the target points are members by observing the reacts of the gradients on them. The accuracy of the active inference attack would be boosted by a significant increase under a global attacker.

Zhang et al. \cite{zhang2020gan} focused on the scenario that the attack is launched by one of the participants and proposed a passive attack using the generative adversarial network (GAN). The attack uses GAN to enrich attack data and increase data diversity that is used to query the target collaborative learning model. The models trained using the new sample-label pairs are vulnerable to membership inference attacks. Yuan et al. \cite{yuan2021beyond} explore record data leakage against NLP in asynchronous distributed learning which would cause imbalanced performance of training across participants. Through eavesdropping on the selection of participants or injecting a single watermark into the victim, they can successfully obtain the privacy records and reveal the identifies of participants.

\subsection{Property Inference}
With the aggregated updates from server, attackers could gradually determine the class representation (i.e. property) of participants' training data. For example,
Hitaj et al. \cite{hitaj2017deep} proposed a GAN-based attack to extract class representation information from honest participants in collaborative learning systems. The attack employs a GAN to generate instances that are visually similar to the samples from a targeted class of a victim participant. In particular, the attack first generates some fake samples from the targeted class that are injected into the training dataset as another class. In this way, the victim participant would reveal privacy information about the targeted class since he has to distinguish between the two classes. Taking the advantage of the knowledge about the targeted class and the density estimation of GAN, the attacker can learn the distribution of the targeted class without accessing the training points of the victim participant directly. The attack is effective against the collaborative learning tasks with convolutional neural networks, even when the parameters are obfuscated via differential privacy techniques.

The GAN-based class representation attack infers only properties of the entire targeted class and assumes that the victim participant owns the entire training points of the targeted class. In contrast, Melis et al. \cite{melis2019exploiting} release the constrained assumptions and proposed property inference attacks to extract unintended information about participants’ training data from the update history. Specifically, at each training iteration, the attacker saves the snapshot of the aggregated update parameters. The difference between the consecutive snapshots is equal to the aggregated updates from all participants. This difference directly reveals privacy information in the training batches of the honest participants during collaborative learning. Melis et al. proposed property attacks in both passive and active modes:
\begin{itemize}
	\item Passive property inference: consider that the attacker has auxiliary data consisting of the data points that have the property of interest and data points that do not have the property. The intuition behind the attack is that the adversary can leverage the snapshots of the global model to generate aggregated updates based on the data with the property and updates based on the data without the property. This produces labeled examples that enable the adversary to train a binary batch property classifier that determines if the observed updates are based on the data with or without the property.
	\item Active property inference: the active attacker can perform a more powerful attack using multi-task learning. The adversary extends his local copy of the collaboratively trained model with an augmented property classifier connected to the last layer. He trains this model to simultaneously perform well on the main task and recognize batch properties.
\end{itemize}

Similar to \cite{hitaj2017deep}, Wang et al. \cite{wang2019beyond,song2020analyzing} proposed GAN-based attacks against collaborative learning systems to target client-level privacy. In the proposed attack, the parameter server is malicious and cannot access the target data. Since GANs could generate conditioned samples, the attacker trains GANs conditioned on the updates from the victim participants, thus it could generate victim-conditioned samples which contain client-level privacy information. They also consider both passive and active modes.
\begin{itemize}
	\item Passive inference: the malicious server is assumed to be honest-but-curious and only analyzes the updates from the participants by training GANs.
	\item Active inference: the active attacker isolates the victim participants from the others, i.e., training GANs on the victim alone by sending a special version of the aggregated model to the victim participants.
\end{itemize}

The above methods require updates information during the training process, i.e., in the white-box mode. Instead, Zhang et al. \cite{zhang2021leakage} suppose the adversary can only black-box access to the global model. Depending on the correlation relationship sensitive attributes with other attributes or labels, they train a sequence of shadow networks and a meta-classifier to learn the distribution of sensitive attributes in a few queries.

\subsection{Sample Inference}
Collaborative learning systems use the gradient sharing framework to avoid data leakage of participants, which is less effective in recent sample inference attacks. Zhu et al. \cite{zhu2019deep} first pointed out that the sharing gradients can leak private training data. They presented an optimization algorithm, deep leakage from gradients (DLG), that can obtain both the training inputs and the labels in just a few iterations. The attack first randomly generates a pair of ``dummy'' inputs and labels and then derives the dummy gradients from the dummy data. The attack optimizes the dummy inputs and labels to minimize the distance between dummy gradients and real gradients. The private training data will be fully revealed by matching the gradients makes the dummy data close to the original ones.

Although DLG works, Zhao et al. \cite{zhao2020idlg} found that it is not able to reliably extract the ground-truth labels or generate good quality training samples. Zhao et al. proposed a simple yet efficient sample inference attack to extract the ground-truth labels from the shared gradients. They demonstrate that the gradient of the classification loss can distinguish correct label from others by derivation. With such observation, the attacker can identify the ground-truth labels based on the shared gradients. Then, the attacker can significantly simplify the DLG attack and extract good-quality training samples.

Later, numerous sample reference attacks are devoted to improve the effectiveness of the revealing training samples and labels \cite{yin2021see, dang2021revealing, jin2021cafe, fulabel,chen2021knowledge}. For example, Yin et al. \cite{yin2021see} present GradInvision to recover a single image from the averaged gradients. In particular, GradInvision first performs label revealing from the gradients of the fully-connected layer and then optimizes random inputs to match the target gradients using fidelity regularization and get a better quality of the reconstructed image. Dang et al. \cite{dang2021revealing} consider that participants compute updates with reasonable small batch size and proposed RLG (Revealing Labels from Gradients) that reconstructs training samples from only the gradient of the last layer. Meanwhile, Chen et al. \cite{chen2021knowledge} and Fu et al. \cite{fulabel} investigate the large-batch data leakage in vertical federated learning and He et al. \cite{he2019model} explore the sample reconstruction in the model parallelism architecture.

The above sample inference attacks maily rely on two components: the euclidean cost function and optimization via LBFGS. Geiping et al. \cite{geiping2020inverting} argue that these choices are not optimal for more realistic architectures and especially arbitrary parameter vectors and proposed to use a cost function based on angles, i.e. cosine similarity. One the one hand, the magnitude measures local optimality of the data point and only captures information about the state of training. On the other hand, the angle quantifies the change in prediction at one data point when taking a gradient step towards another.

\section{Privacy Defenses}\label{sec:privacy_defense}
Inspired by the privacy attacks, a large amount of privacy defenses are proposed to protect the training samples from being inferred. According to the commonly used privacy-preserving techniques, we can classify existing privacy defenses into three categories: differentially private, privacy-preserving and cryptographic privacy-preserving collaborative learning. We elaborate state-of-the-art privacy defenses as below.

\subsection{Differentially Private Collaborative Learning}
To mitigate membership inference attacks in collaborative training systems, one promising solution is Differential Privacy (DP), which is a rigorous mathematical framework to preserve the privacy of individual data records in a database when the aggregated information about this database is shared among untrusted parties \cite{dwork2006our,dwork2010boosting}. A number of studies have applied DP to enhance the privacy of DL training in different environments \cite{chaudhuri2011differentially,abadi2016deep,zhang2018improving,li2018differentially,yu2019differentially,jayaraman19evaluatiing}. Most existing DP-SGD algorithms adopt additive noise mechanisms by adding random noise to the estimates in every training iteration. There exists a trade-off between privacy and usability, determined by the noise scale added during training: adding too much noise can meet the privacy requirements, at the cost of the drop in model accuracy. As a result, it is critical to identify the minimal amount of noise that can provide desired privacy protection as well as maintain acceptable model performance.

Two common approaches were devised to optimize the DP mechanisms and balance the privacy-usability trade-off. The first one is to carefully restrict the sensitivity of randomized mechanisms. For example, Abadi et al. \cite{abadi2016deep} bounded the influence of training samples on gradients by clipping each gradient in $l_2$ norm below a given threshold. Yu et al. \cite{yu2019differentially} optimized the model accuracy by adding decay noise to the gradients over the training time since the learned models converge iteratively. The second approach is to precisely track the accumulated privacy cost of the training process using composition techniques such as the strong composition theorem \cite{dwork2010boosting} and moments account (MA) \cite{abadi2016deep, bhowmick2018protection,hynes2018efficient,kang2019weighted}. In the following, we first illustrate exiting frequently used DP techniques and then summarize differentially private solutions for collaborative learning systems.

\subsubsection{DP Techniques}
For any two neighboring datasets that only differ in one single record, a randomized mechanism $\mathcal{M}$ is differentially private if its outputs are almost the same on the two datasets. The formal definition of DP is illustrated as follows.
\begin{definition}(($\epsilon, \delta$)-DP)
    A randomized mechanism $\mathcal{M}: D \rightarrow R$ with domain $D$ and range $R$ satisfies ($\epsilon, \delta$)-DP if for any two neighboring datasets $D_1, D_2$ and any subset of outputs $S \subseteq  R$, the following property is held:
    \begin{equation}
        Pr[\mathcal{M}(D_1) \in S] \leq e^{\epsilon}Pr[\mathcal{M}(D_2) \in S] + \delta.
    \end{equation}
\end{definition}
$\mathcal{M}$ can satisfy DP is restricted by two parameters: $\epsilon$ and $\delta$. $\epsilon$ is the privacy budget to limit the privacy loss of individual records. $\delta$ is a relaxation parameter that allows the privacy budget of $\mathcal{M}$ to exceed $\epsilon$ with probability $\delta$. It has been proven that differential privacy satisfies a composition property: when  with privacy budgets $\epsilon_1$ and $\epsilon_2$ are performed on the same data, the privacy budget of the combined of the two mechanisms equals to the sum of the two privacy budgets, i.e., $\epsilon_1 + \epsilon_2$.

\bheading{Relaxed Definitions.} Due to the composition property, composing multiple differentially private mechanisms leads to a linear increase in the privacy budget and the scale of the corresponding noise increases to maintain a fixed total privacy budget. To get a better privacy-usability trade off, multiple DP techniques reduce the this linear composition bound at the cost of slightly increasing the failure probability. There are two commonly used relaxations of differential privacy that  exhibits better accuracy than ($\epsilon, \delta$)-DP, which use different versions of divergences to calculate the distributional difference between the outputs of $\mathcal{M}$ in adjacent datasets: Concentrated Differential Privacy (CDP) and R\'enyi Differential Privacy (RDP). CDP uses the sub-Gaussian divergence to restrict the mean and standard deviation of the privacy loss variable. It get better accuracy and any $\epsilon$-DP algorithm satisfies $(\epsilon \cdot (e^{\epsilon}-1)/2, \epsilon)$-CDP. R\'enyi DP (RDP) \cite{mironov2017renyi} is a natural relaxation of DP based on the R\'enyi divergence and allows tighter analysis of tracking cumulative privacy loss. The instantiation of RDP is MA, which keeps track of a cumulative bound on the moments of the privacy loss.

\subsubsection{DP-SGD for Collaborative Learning}
For single-party learning, there are two common candidates for where to add random noise: the objective function \cite{chaudhuri2011differentially,phan2016differential} and gradients \cite{abadi2016deep, yu2019differentially}. For the first approach, Chaudhuri et al. \cite{chaudhuri2011differentially} perturb the objective function before optimizing over classifiers and show that the objective perturbation is DP if certain convexity and differentiability criteria hold. Phan et al. \cite{phan2016differential} attempt to use the objective perturbation by replacing the non-convex function with a convex polynomial function. To this end, Phan et al. \cite{phan2016differential} design a convex polynomial function to approximate the non-convex one, which however would change the learning protocol and even sacrifice the model performance. On the other hand, adding random noise to the gradients is a simpler and popular approach in single-party learning. For example, Abadi et al. \cite{abadi2016deep} bound the influence of training samples on gradients by clipping each gradient in $l_2$ norm below a given threshold to restrict the sensitivity of randomized mechanisms.
Yu et al. \cite{yu2019differentially} focus on differentially private model publishing and optimize the model accuracy by adding decay noise to the gradients over the training time since the learned models converge iteratively.

Another way to improve the model usability lies in precisely tracking the cumulative privacy cost of the training process. For example, Shokri et al. \cite{shokri2015privacy} and Wei et al. \cite{wei2020federated} composed the additive noise mechanisms using the advanced composition theorem \cite{dwork2010boosting}, leading to a linear increase in the privacy budget. Some DP-SGD methods \cite{abadi2016deep, bhowmick2018protection,hynes2018efficient,kang2019weighted} use MA to reduce the added noise during the training process. Other algorithms \cite{park2017dp,jayaraman2018distributed,yu2019differentially} were designed to improve the model usability using (zero) concentrated DP \cite{dwork2016concentrated}.

Some works~\cite{shokri2015privacy,bhowmick2018protection,hynes2018efficient,jayaraman2018distributed,kang2019weighted, han2021accurate,wei2021user,wei2021gradient,sun2021pain,mao2021romoa,xiong2021privacy} applied the DP techniques from the standalone mode to the distributed systems to preserve the privacy of the training data for each agent. For example, Shokri et al. \cite{shokri2015privacy} proposed a privacy-preserving distributed learning algorithm by adding Laplacian noise to each agent's gradients to prevent indirect leakage. Kang et al. \cite{kang2019weighted} adopted weighted aggregation instead of simply averaging to reduce the negative impact caused by uneven data scale in collaborative learning systems.

In terms of the accumulated privacy loss, Kang et al. \cite{kang2019weighted} employed MA to track the overall privacy cost of the collaborative training process. Wei et al. \cite{wei2020federated} and Wei \cite{wei2021user} perturbed agents' trained parameters locally by adding Gaussian noise before uploading them to the server for aggregation and bounded the sensitivity of the Gaussian mechanism by clipping in federated learning systems. Shokri et al. \cite{shokri2015privacy} and Wei et al. \cite{wei2020federated} composed the additive noise mechanisms using the strong composition theorem \cite{dwork2010boosting}, leading to a linear increase in the privacy budget. In order to reduce aggregated noise in local updates, Han et al. \cite{han2021accurate}dynamically adjust batch size and noise level according to the rate of critical input data and the sensitivity estimation.

\subsection{Cryptographic Privacy-preserving Collaborative Learning}
Although DP techniques are widely used in collaborative learning as its clear theory and concise algorithm, they are designed against membership inference attacks and hard to defend property and sample inference attacks. Additionally, the noise added to the updates would reduce the performance of the trained models, especially when the participants are extremely sensitive to privacy leakage. Due to the drawbacks of DP techniques, several privacy-preserving collaborative learning methods are proposed using cryptographic tools as we elaborate below.

\bheading{Collaborative Learning with Homomorphic Encryption.}
Homomorphic Encryption(HE) permits users to perform arithmetic operations directly on ciphertext, which is equivalent to executing the same operations on the corresponding plaintext. Since HE techniques only require participants to share encrypted data, they can provide cryptographic privacy protection in collaborative learning scenarios. There are two types of HE schemes: Fully Homomorphic Encryption(FHE) and Partially Homomorphic Encryption(PHE). FHE allows addition and multiplication on encrypted data, while PHE only support one of them. Correspondingly, FHE is much more computationally expensive than PHE.
Several privacy-preserving collaborative learning approaches have been proposed to use PHE to ensure the privacy of individual model updates \cite{aono2016scalable,2017Privacy, 2020Privacy, 2021GALA,zhang2020batchcrypt}. For example, Aono \cite{aono2016scalable}, Phong \cite{2017Privacy}, and PPFDL \cite{2020Privacy} perform the addition operation over encrypted updates to protect the privacy of the updates during the aggregation process.

To reduce the cost of homomorphic linear computation, Zhang et al. \cite{2021GALA} considers homomorphic linear computation as a sequence addition operations of addition, multiplication, and permutation and then greedy chooses the least expensive operation for every computation step. Froelicher et al. proposed SPINDLE \cite{froelicher2021scalable} that preserves data and model confidentiality and enables the execution of a cooperative gradient-descent and the evaluation of the obtained model even when there are colluding participants. Stripelis et al. \cite{stripelis2021secure} proposed a secure federated learning framework FL using FHE techniques to protect training data and the shared updates.
\cite{stripelis2021secure}
However, HE has some limitations. For example, the memory and arithmetic cost of encrypted data is much higher than that of the original plaintext. And HE has to use polynomial approximations to handle common non-linear operations in collaborative learning systems.

\bheading{Collaborative Learning with Secure Muti-Party Computation.}
Another widely applied cryptographic method is secure multi-party computation (SMC), which allows mutual distrust participants to jointly compute a function over their inputs and preserve the privacy of inputs \cite{bonawitz2017practical, bell2020secure, li2020privacy, li2020toward}. Bonawitz et al. \cite{bonawitz2017practical} proposed a communication-efficient, failure-robust secure aggregation of high dimensional model updates without learning each participant's sensitive information with SMC, which can defense both passive and active adversaries. Li et al. \cite{li2020privacy} proposed a privacy-preserving collaborative learning framework based on the chained SMC technique. With such framework, adversaries can not obtain any privacy of participants as the output of a single participant is dissimulated with its prior. Comparing to HE, SMC has less computation cost and communication overhead, but it is still not applicable to large-scale collaborative learning, especially collaborative learning systems with thousands of participants.

\subsection{Practical Privacy-preserving Collaborative Learning}
\begin{figure}[t]
	\centering
	\includegraphics[width=0.9\columnwidth]{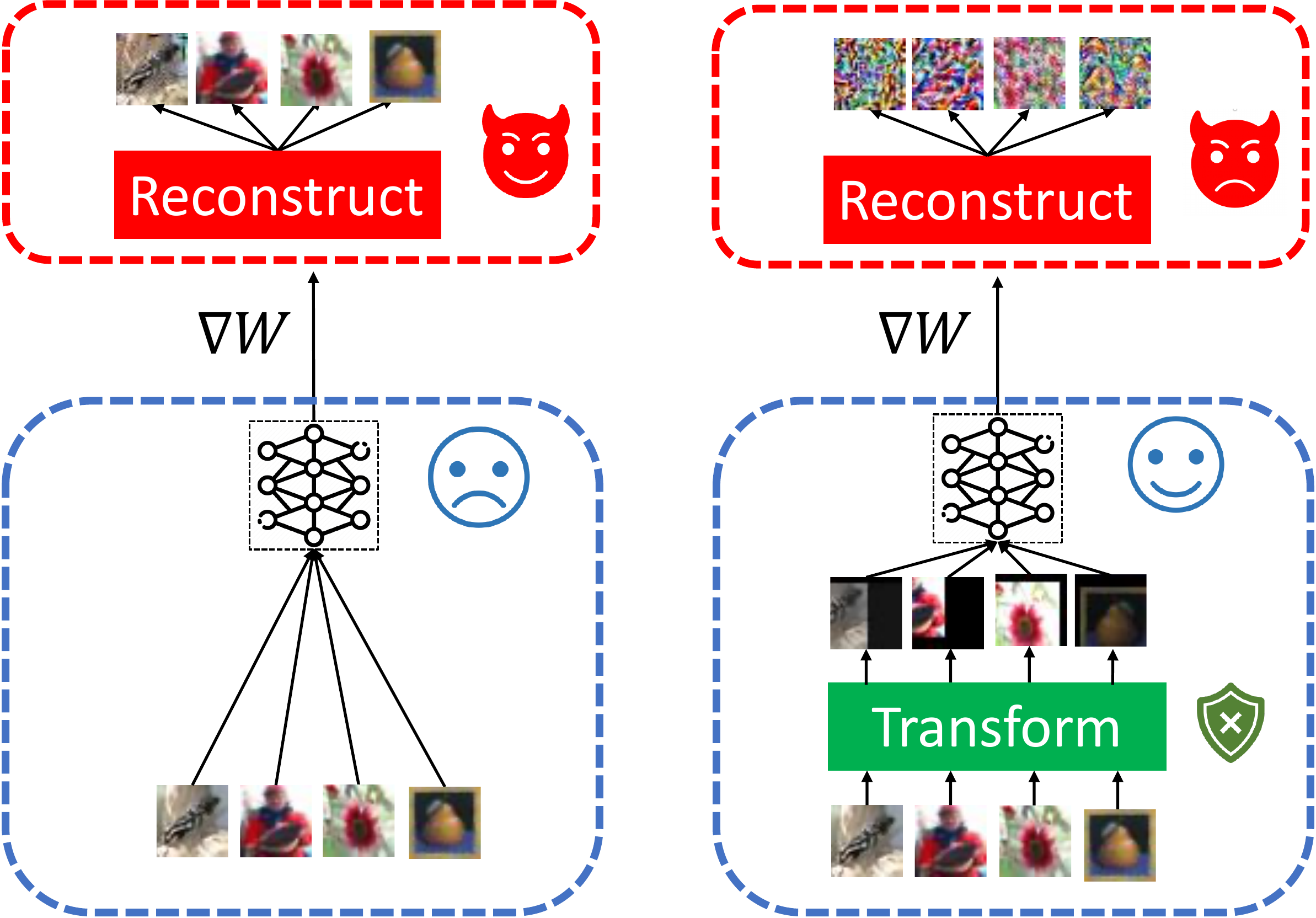}
	\caption{Automatic transformation search against deep leakage from gradients \cite{gao2021privacy}.}
	\label{fig:privacy-defense}
\end{figure}

Besides the above privacy defenses whose security can be theoretically guaranteed, a large amounts of privacy-preserving collaborative learning methods are proposed to protect the privacy of participants in real collaborative learning scenarios. Similar to integrity defenses, these privacy defenses also focus on processing training data or model updates to experimentally guarantee privacy information against inference attacks. Figure ~\ref{fig:privacy-defense} illustrates a privacy-preserving collaborative learning method \cite{gao2021privacy} using automatic transformation search against deep leakage from gradients. The method transforms original local data samples into related samples for disabling sample inference attacks by searching specific transformation.

Such approach is much more efficient than the defenses with cryptographic techniques to thwart inference attacks. Zhao et al. \cite{zhao2020privatedl} proposed a framework that transfers sensitive samples to public ones with privacy protection, based on which the participants can collaboratively update their local models with noise‐preserving labels. Fan et al. \cite{fan2020rethinking} designed a secret polarization network for each participant to produce secret losses and calculate the gradients. Huang et al. \cite{huang2021evaluating}  proposed to combine existing sample inference defenses in an appropriate manner to enhance the protection performance.
\section{Hybrid Defenses and Beyond}
\subsection{Hybrid Defenses}
Existing investigations \cite{naseri2020toward} have shown that defenses against one type of attacks cannot be directly applied to the other type of attacks. Thus, besides the defenses that aim to prevent one type of threats, several methods \cite{ma2022privacy,grama2020robust,qi2021privacy,liu2021privacy,lyu2021dp,dong2021flod, domingo2021secure} are proposed to defend both integrity and privacy attacks and build robust and privacy-preserving collaborative learning systems. Theses hybrid defenses mainly take advantage of techniques against both integrity and privacy attacks. We elaborate state-of-the-art hybrid defenses as below.

One main design strategy of hybrid defenses \cite{ma2022privacy,grama2020robust,liu2021privacy} are to merge existing defenses for integrity and privacy defenses together to achieve secure collaborative learning systems. For example, Ma et al \cite{ma2022privacy} utilize an existing Byzantine-robust federated learning algorithm and distributed Paillier encryption and zero-knowledge proof to guarantee privacy and filter out anomaly parameters from Byzantine participants. Qi et al \cite{qi2021privacy} achieve hybrid defense using blockchain and differential privacy techniques.

Some hybrid defenses leverage homomorphic encryption techniques that provide both confidentiality and computability over encrypted data. For example, Liu et al \cite{liu2021privacy} proposed a homomorphic encryption scheme that enables privacy protection and provides the parameter server a channel to punish poisoners under ciphertext. Dong et al \cite{dong2021flod} employ two non-colluding servers and proposed an oblivious defender for private Byzantine-robust federated learning using additive homomorphic encryption and secure two-party computation  primitives. However, homomorphic encryption based defenses requires a large amount of computations resources. Domingo et al \cite{domingo2021secure} offer privacy to the participants as well as robustness against Byzantine and poisoning attacks via unlinkable anonymity, which can detect bad model updates while reducing the computational overhead compared to homomorphic encryption based defenses.

\subsection{Collaborative Adversarial Training}
Deep models are vulnerable to adversarial examples that are maliciously constructed to mislead the models to output wrong predictions but visually indistinguishable from normal samples \cite{goodfellow2014explaining,carlini2017towards,yuan2019adversarial,xiang2021local}. Adversarial training \cite{shafahi2019adversarial,tramer2017ensemble,wong2020fast} is one of the most effective approaches to defend deep models against
adversarial examples and enhance their robustness. Its main idea is to augment training data with existing adversarial example generation methods during the training process. Thus, the adversarially trained models are more robust against adversarial examples during the inference process. Numerous of adversarial training methods \cite{bai2021recent} have been proposed for standalone training systems. For example, Shafahi et al \cite{shafahi2019adversarial} proposed an efficient adversarial training algorithm that recycles the gradient information computed at each iteration to eliminate the overhead cost of generating adversarial examples. Wong et al \cite{wong2020fast} propose to utilize Fast Gradient Sign Method (FGSM) \cite{goodfellow2014explaining} during the adversarial training process. They introduce random initialization points to improve the effectiveness the projected gradient descent based training.

Although standalone adversarial training achieves great success, challenges rise when extending these standalone algorithms to collaborative learning systems. One main challenge is that participants in collaborative systems only have limited training data and computational resources and cannot support the data-hungry and costly adversarial training. To this end, several adversarial training algorithms \cite{hong2021federated,zhou2020adversarially,shah2021adversarial} have been proposed specifically for collaborative learning systems. For instance, Hong et al \cite{hong2021federated}  proposed an effective propagation methods that transfers adversarial robustness from high-resource participants that can afford adversarial training to low-resource participants. Zhou et al \cite{zhou2020adversarially} conduct collaborative adversarial training by composing the aggregation error of the parameter server(s) into bias and variance and using the bias-variance adversarial examples to improve model robustness. Shah et al \cite{shah2021adversarial} consider communication constrained federated learning environments and proposed an dynamic adversarial training methods to improve both adversarial robustness and model convergence speed.
\section{Open Problem}\label{sec:open_problem}
Although extensive research has been proposed to address the integrity and privacy threats in collaborative learning, some interesting and important issues remain to be fully explored. Here, we present several open problems and potential research directions below to stimulate further research:

\bheading{Non-IID or Noisy Scenarios in Byzantine Attacks and Defenses.} Byzantine attacks and defenses is an arms race between attackers and defenders: attackers intend to design malicious updates that are indistinguishable from normal ones, while defenders try to identify potential Byzantine updates and ensure the integrity of the trained models. Most of existing Byzantine resilient algorithms only consider IID training scenarios, where the training datasets of benign participants are IID. However, the training datasets are Non-IID in most real cases because the quality and distribution of each training dataset is different. Thus, it is more difficult for defenders to distinguish benign and malicious updates. For example, a malicious participant can disguise as a node with poor training data quality and generate updates that are indistinguishable from normal ones but fatal to model integrity. Although several works \cite{xie2019zeno,cao2021fltrust} attempt to propose Byzantine resilient aggregation rules in Non-IID scenarios, they fail to defend advance Byzantine attacks or only consider few types of Non-IID scenarios \cite{cao2021fltrust}.

\bheading{Certified Backdoor Defenses.} Existing backdoor defenses for collaborative learning mainly focus on empirically identifying or removing backdoors. Such defenses work well for existing backdoor attacks but can hardly identify or remove new advance attacks. Therefore, certified backdoor defenses for collaborative learning are urgently needed and provide provable protection against backdoor attacks. Unfortunately, most of existing certified backdoor defenses \cite{weber2020rab,wang2020certifying} are proposed for standalone machine learning systems and few of them \cite{xie2021crfl} are designed for collaborative learning.

\bheading{Privacy-performance Tradeoff in Differential Privacy.} Differential privacy techniques require adding noise onto the updates/models to defend membership inference attacks. Although several relaxation methods have been proposed to reduce the scale of noise, the performance is still unsatisfactory, especially when the parameters of the trained neural networks are large \cite{guo2021topology,wei2021user}. One possible research direction is to utilize the system features of collaborative learning systems to reach a better privacy-performance tradeoff.

\bheading{Basis Datasets in Property Inference Attacks.} Several attacks \cite{hitaj2017deep,melis2019exploiting} utilize local datasets to infer the property of other participants. Such local datasets are assumed to have the same distribution with victim participants and critical to the inference attacks. However, such IID datasets limit the threat of these attacks because adversaries may not know the distribution of the training datasets of victim participants. Thus, how to conduct property inference attacks with basis datasets is worth of in-depth study.

\bheading{Performance Improvement in Sample Inference Defenses.} Samples inference defenses \cite{gao2021privacy,huang2021evaluating} can protect training samples from being inferred by existing attacks. However, some defenses such as adding noise or pruning parameters would harm the performance of the collaboratively trained models. Thus, it is necessary to design new defenses that can enhance both privacy and performance of collaborative learning.

\section{Conclusion}\label{sec:conclusion}
In this survey, we systematically introduce state-of-the-art integrity and privacy threats in collaborative learning systems, which mainly include byzantine and backdoor attacks as well as three kinds of data inference attacks. And we also detailed describe corresponding defensive policies, including model and data based inspection against integrity attacks and differential privacy and encryption techniques against privacy attacks. The tradeoff between security and performance is the crucial point of new defense schemes.
Additionally, we discuss a number of open problems of current defense methods, hoping it could help researchers identify and solve issues more quickly in the area of robust and privacy- preserving collaborative learning.

\bibliographystyle{./IEEEtran}
\bibliography{ref}

\end{document}